\documentclass[aps,preprint]{revtex4}%

\usepackage{graphicx}
\usepackage{epsfig}

\newcommand{\be}{\begin{equation}}
\newcommand{\ee}{\end{equation}}
\usepackage{amsfonts}
\usepackage{amsmath}
\usepackage{amssymb}
\usepackage{graphicx}%
\begin{document}
\title{The Physics Behind High-Temperature Superconducting
Cuprates: The ``Plain Vanilla'' Version Of RVB}
\author{
P. W. Anderson$^{(1)}$, P. A. Lee$^{(2)}$, M. Randeria$^{(3)}$,
T. M. Rice$^{(4)}$, N. Trivedi$^{(3)}$ and F. C. Zhang$^{(5)}$
}
\affiliation{
(1) Department of Physics, Princeton University, Princeton, NJ 08544 \\
(2) Department of Physics, Massachusetts Institute of Technology, 
Cambridge, MA 02139 \\
(3) Tata Institute of Fundamental Research, Mumbai 400005, India \\
(4) Theoretische Physics, ETH, H\"onggerberg, CH 8093 Zurich, Switzerland \\
(5) Department of Physics, The University of Hong Kong, Hong Kong, and \\
Department of Physics, University of Cincinnati, Cincinnati, OH 45221
}

\begin{abstract}
One of the first theoretical proposals for understanding high temperature
superconductivity in the cuprates was Anderson's RVB theory using a
Gutzwiller projected BCS wave function as an approximate ground state.
Recent work by Paramekanti, Randeria and Trivedi has shown that this 
variational approach
gives a semi-quantitative understanding of the doping dependences of a
variety of experimental observables in the superconducting state of the
cuprates.  In this paper we revisit these issues using the ``renormalized
mean field theory'' of Zhang, Gros, Rice and Shiba based on the 
Gutzwiller approximation
in which the kinetic and superexchange energies are renormalized by different
doping-dependent factors $g_{t}$ and $g_{S}$ respectively. We point out a
number of consequences of this early mean field theory for experimental
measurements which were not available when it was first explored, and observe
that it is able to explain the existence of the pseudogap, properties of
nodal quasiparticles and  approximate spin-charge separation, the latter
leading to large renormalizations of the Drude weight and superfluid density.
We use the Lee-Wen theory of the phase transition as caused by thermal
excitation of nodal quasiparticles, and also obtain a number of further
experimental confirmations. Finally, we remark that superexchange, and not
phonons, are responsible for d-wave superconductivity in the cuprates.

\end{abstract}
\maketitle

\noindent
{\bf Introduction}

The resonating valence bond (RVB) liquid was suggested in 1973 by 
Anderson and Fazekas (Anderson, 1973; Fazekas and Anderson, 1974)
as a possible quantum state for antiferromagnetically coupled 
$S=1/2$ spins in a low dimensions. Their ideas were based on 
numerical estimates of the ground state energy.
Instead of orienting the atomic magnets on separate, 
oppositely-directed sublattices, in the liquid they were supposed 
to form singlet ``valence bonds'' in pairs, and regain some of
the lost antiferromagnetic exchange energy by resonating 
quantum-mechanically among many different pairing configurations. 
Such states form the basis of Pauling's early theories of
aromatic molecules such as benzene (as well as of his unsuccessful 
theories of metals), and are a fair description of Bethe's (1931) 
antiferromagnetic linear chain.  
The $S = 1/2$ antiferromagnetic Heisenberg model arises naturally
in Mott insulators. Unlike conventional band insulators,
Mott insulators have an odd number of electrons per unit cell and are
insulating by virtue of the strong Coulomb repulsion between two
electrons on the same site. Virtual hopping favors anti-parallel
spin alignment, leading to antiferromagnetic exchange coupling $J$
between the spins (Anderson, 1959).
In the RVB picture, $S=1/2$ is important because 
strong quantum fluctuations favor singlet formation 
rather than the classically ordered N\'{e}el state.

In 1986 the high $T_c$ cuprates were discovered (Bednorz and Muller, 1986),
and it was soon realized (Anderson, 1987a) that the operative
element in their electronic structures was the square planar CuO$_2$ lattice.
In the ``undoped'' condition, where the Cu is stoichiometrically Cu$^{++}$,
the CuO$_2$ plane is just such an antiferromagnetically coupled
Mott insulator. In many instances these planes are weakly coupled to 
each other.
Anderson (1987a, 1987b), in response to this discovery, showed that an RVB
state could be formally generated as a Gutzwiller projection of a BCS pair
superconducting state.  This is a much more convenient and
suggestive representation than those based on atomic spins, and it
immediately makes a connection with superconductivity.

The method of Gutzwiller (1963) was initially proposed as a theory of 
magnetic metals, in conjunction
with the Hubbard model. His proposal was to take into account the 
strong local Coulomb repulsion
of the electrons by taking a simple band Fermi sea state and simply 
removing, by projection,
all (or, in the early version, a fraction) of the components in it 
which have two electrons on the
same site. When one projects a half-filled band in this way the 
result is to leave only singly-occupied
sites with spins.  The new idea is to project a BCS paired 
superconducting state; then the spins are
paired up in singlet pairs to make a liquid of pair ``bonds''; see Fig.~1.

\begin{figure}
\begin{center}
\vskip-2mm
\hspace*{0mm}
\epsfig{file=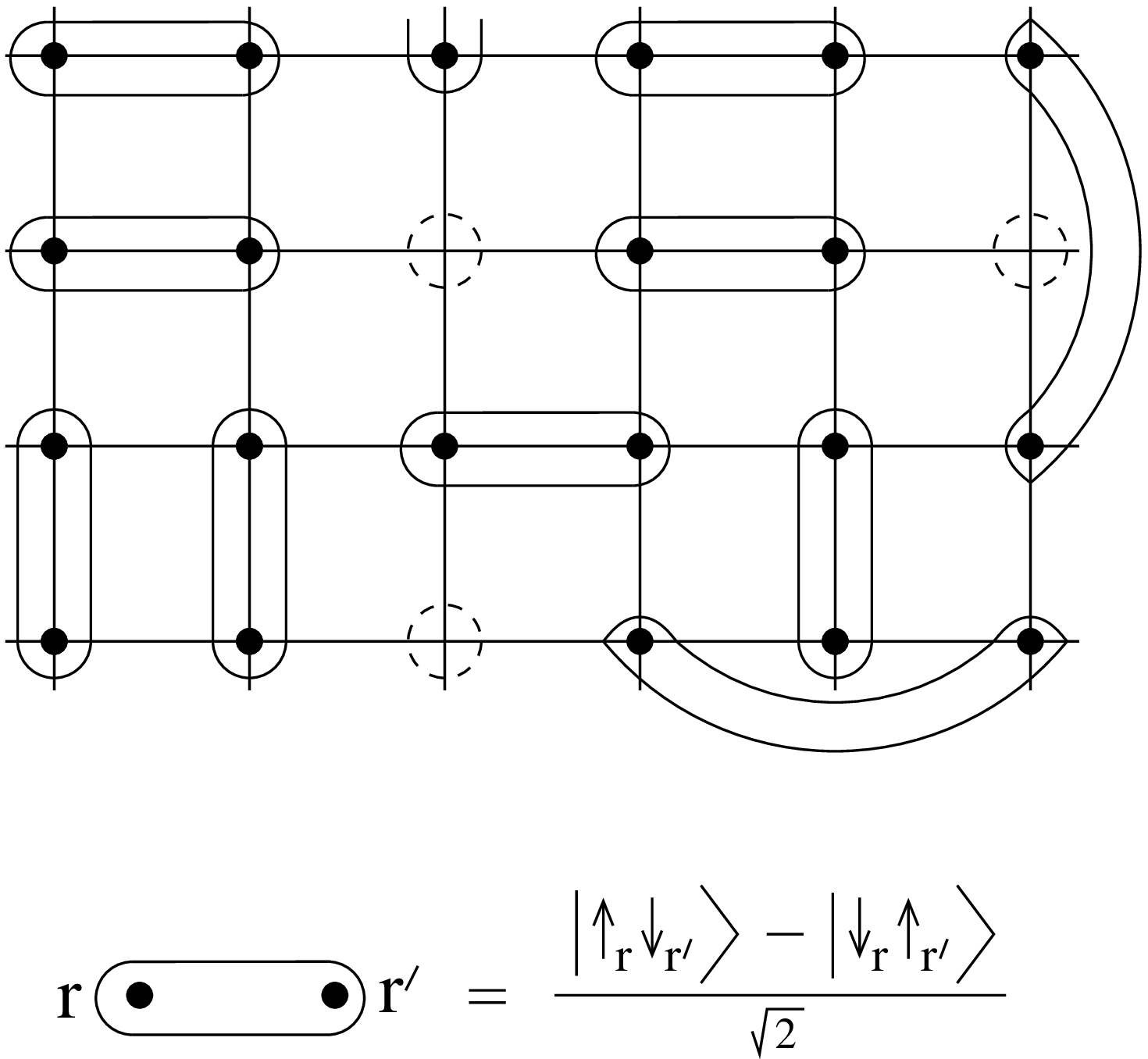,height=4.0in,width=4.0in,angle=0}
\vskip-6mm
\caption{
Snapshot of a resonating valence bond (RVB) configuration showing
singlet pairs of electrons and, in addition, a fraction $x$ of doped holes.
The many-body ground state wavefunction is a linear superposition of such configurations
with the spatial dependence of the singlet pairing amplitudes determined
by the function $\varphi(r - r^\prime)$ defined in eq.~(4).
}
\label{fig:rvb}
\vskip-6mm
\end{center}
\end{figure}

But of course, with exactly one spin at every site,
this state is a Mott insulator, not a metal. Such an 
RVB liquid state is of rare occurrence
in real Mott insulators, which usually exhibit either 
antiferromagnetic long range order
as in the cuprates, or possibly have ordered ``frozen'' arrays of 
bonds. i.e., valence bond
crystals rather than liquids.  However, the importance of the RVB 
liquid was the suggestion that
as one doped this state with added electrons or holes, the
resulting metal would be a high Tc superconductor, retaining the 
singlet pairs but allowing them to
carry charge and support supercurrents. The motivation for the 
pairing would be the antiferromagnetic
superexchange of the original Mott insulator.

For over a decade and a half a number of theorists have been trying 
to implement this suggestion
along a bewildering variety of routes. One main avenue has resulted 
from the proposal by several
authors (Kotliar and Liu, 1988; Suzumura {\it et al.}, 1988; Gros, 1988;
Yokoyama and Shiba, 1988; Affleck {\it et al.} 1988; Zhang{\it et al.} 1988),
that Anderson's original s-wave BCS be replaced by an exotic, d-wave state. 
The d-wave approach in the early days was quantitatively carried
through by Gros (1989) using variational Monte Carlo methods 
and by Zhang, Gros, Rice and Shiba (1988) on a simplified model,
and using very rough approximation methods.
Recently the Gutzwiller-RVB wavefunction approach was revived
by Paramekanti, Randeria, and Trivedi (2001; 2003) who used
careful numerical methods to calculate many quantities of direct
experimental relevance. Their results turn out to correspond remarkably
well to the experimental phenomena observed in the cuprates across a very broad
spectrum of types of data, a spectrum that was simply  not
available in 1987-88 when the original work was done. It may be
because of this absence of data at the time that the
original paper was for so long not followed up.

All of this work relies on one basic assumption, an
assumption which has gone unquestioned among a large
fraction of those theorists concerned with this problem,
from the beginning. This is the assumption that the physics
of these materials is dominated by the strong repulsive
interactions of a single non-degenerate band of electrons on
the CuO$_2$ planes, and is specifically not at all similar to that
of the conventional BCS superconductors. In the latter the
direct electron interactions are heavily screened, and the
lattice vibrations play the dominant role. We feel that the
demonstration of d-wave superconductivity in particular
makes phonons as major players difficult to support, even
though there are some notable physicists, such as Mott,
Friedel, Muller, and Abrikosov, who disagree.
The phonon mechanisms are local in space,
extended in time, making the dynamic screening mechanism
emphasized by Schrieffer and Anderson relevant and leading to
s-wave pairing (Schrieffer, 1964). This mechanism works better the more
electrons there are per unit cell, and fails for monovalent
metals. D-wave pairing, on the other hand, is essentially
non-local in space and deals with strong repulsions by
conventional space avoidance, as suggested by Anderson and
Morel (1961) and by Kohn and Luttinger (1965). Phonon interactions,
especially via optical phonons, are local and cannot easily
lead to higher angular momentum pairing. 

It has been argued that 
certain specific phonons in the presence of strong correlation can 
enhance $d$-wave pairing (Shen {\it et al}., 2002).  However, such 
couplings are reduced for small doping by the renormalization factor 
$g_t^2$ as discussed later in the article.
Even more cogent is the fact
that, as we shall see, the attractive potential for d-wave pairing
is more than adequate without phonons, and even if they
contribute positively to it the effect will be  minor. (It
has been argued that in some cases the contribution is
negative (Anderson 2002)).

Furthermore, it is now known that the energy gap in high $T_c$ 
superconductors is much larger than predicted by BCS theory, and can 
reach a value of order 50~meV.  This is comparable to or exceeds 
typical phonon frequencies, making it obvious that a phonon cannot be 
the key player.

We prefer not to further burden this discussion with the
equally strong chemical, angle resolved photoemission
spectra and optical evidence for using only a single  band;
this subject is treated in, for instance, the paper by Zhang
and Rice (1988), or in Anderson's book (1997).

These considerations suggested the use of models where the
strong repulsive correlations are emphasized, specifically
the Hubbard model, which takes as the only interaction a
strong on-site repulsion. 
The Hubbard model can be
transformed by a perturbative canonical transformation (Kohn 1964) into
a block-diagonal form in which double occupancy is excluded,
and replaced by an exchange interaction between neighboring
sites as pointed out early on by Gros, Joynt and Rice (1986).
This procedure converges well for sufficiently strong on-site
interaction $U$, but presumably fails at the  critical
$U$ for the Mott transition; the singly-occupied ``undoped''
case is unquestionably a Mott insulator in the cuprates and
this transformation ipso facto works. The further simplified
$t-J$ model is often used; for refined calculations it has
been argued (Paramekanti {\it et al.}, 2001; 2003) that this simplification
may be too great, but for the semi-quantitative purposes of this
article we will at least think in terms of that model.

The Mott insulator based theory for the cuprates has been
expressed in a variety of forms other than straightforward
Gutzwiller projection and we do not claim any great overall
superiority for our method. Early on, Baskaran, Zou and
Anderson (1987) (see also, Anderson (1987b) and Zou and Anderson (1988))
introduced the ideas of spin-charge separation
(see also Kivelson, Rokhsar and Sethna, 1987) and of slave bosons
and gauge fields introduced to implement the Gutzwiller constraint, and
quite a number of authors (Ruckenstein {\it et al.}, 1987, Weng {\it 
et al.}, 1996)
have followed this direction,
most notably Baskaran (Anderson, Baskaran, Zou and Hsu, 1987; 
Baskaran and Anderson, 1988),
Fukuyama (Suzumura {\it et al.}, 1988; Fukuyama, 1992)
Kotliar (Kotliar and Liu, 1988), Ioffe and Larkin (1989) and a series 
of publications
by P. A. Lee and co-workers (Nagaosa and Lee, 1990; Wen and Lee, 1996).
A related method is the Schwinger boson, slave
fermion technique which has been discussed by a number  of
authors (Weigmann, 1988; Shraiman and Siggia, 1989; Lee, 1989).
Undoubtedly for discussions of the precise
nature of the phase transition and of the complicated mix of
phenomena such as the pseudogap regime which occur above
$T_c$ these theories will be essential, but we here focus  on
properties of the ground state and of low-lying excitations,
which  by good fortune includes the basic physics of $T_c$.
We feel that what we can calculate indicates the correctness
of  the fundamental Mott-based picture in such a way as to
support the further effort needed to work out these
theories.

\noindent
{\bf The Method}

Starting from the Hubbard Hamiltonian (which may be
generalized in various ways without affecting the following
arguments) %
\be
H=T+U%
{\displaystyle\sum_{i}}
n_{i\uparrow}n_{i\downarrow}%
\ee
where $T$ is the kinetic energy. We suppose that there is  a
canonical transformation $e^{iS}$ which eliminates $U$  from
the     block     which    contains    no    states     with
$n_{i\uparrow}+n_{i\downarrow}=2$,  and   which   presumably
contains all the low-lying eigenstates and thus the ground state;
there  are no matrix elements of the transformed Hamiltonian
connecting these to doubly-occupied states. Thus%

\be
e^{iS}He^{-iS}=H_{t-
J}=PTP+J\sum_{ij}\mathbf{S}_{i}\cdot\mathbf{S}_{j}%
\ee
Here $P=\prod_i\left(  1-n_{i\uparrow}n_{i\downarrow}\right)$
is the Gutzwiller projection operator, which projects out
double   occupancy.  The  kinetic  energy  $T$  is  actually
modified  to  include a 3-site hopping term, which  we  will
neglect here, realizing that our Fermi surface and velocity
are  heuristically adjusted in any case. The low-lying eigenstates
of this Hamiltonian are necessarily of the form $P\left\vert
\Phi\right\rangle $, where
$\left\vert  \Phi\right\rangle $  is  a  completely  general
state of the appropriate number of electrons in the band.
Thus Gutzwiller projection is necessary if one is to use the
canonical transformation to eliminate $U$.

We make the fundamental assumption that the correct $\left\vert
\Phi\right\rangle $ may be approximated by a general product
wave function of Hartree-Fock-BCS type, so that%
\be
P \left\vert \Phi\right\rangle = P \prod_{\vec k}\left( u_{\vec
k}+v_{\vec k}c_{\vec k\uparrow}%
^{\dag}c_{-\vec    k\downarrow}^{\dag}\right)     \left\vert
0\right\rangle.
\ee
In fact, one can simply rewrite $P \left\vert \Phi\right\rangle$
for a fixed number of electrons ($N$) as
\be
P \left\vert \Phi\right\rangle = P
\left[\sum_{{\vec r},{\vec r}^\prime} \varphi({\vec r} - {\vec r}^\prime)
c_{{\vec r}\uparrow}^{\dag}c_{{\vec r}^\prime\downarrow}^{\dag}\right]^{N/2}
\left\vert 0\right\rangle,
\ee
where $\varphi({\vec r} - {\vec r}^\prime)$ is the Fourier transform of
$v_{\vec k}/u_{\vec k}$.
This real space wavefunction may be visualized in terms of a linear 
superposition
of configurations consisting of singlet pairs and vacancies with no
double occupancy. Each valence bond is the snapshot of a preformed pair of
electrons, while the vacancies correspond to doped holes; see Fig.~1.

In the conventional theory of metals, the Hartree-Fock BCS ansatz
turns out to be justifiable as the first step in a
perturbation series which preserves many of the properties
of the non-interacting particle model, relying on adiabatic
continuation arguments in a qualitative way. We see no
reason why it cannot be equally effective in this case.
We emphasize that we are {\em not} approximating the actual wave-
function $e^{iS}P\left\vert \Phi\right\rangle $ as a product
function, but the function to be  projected, $\left\vert
\Phi\right\rangle $, and we are searching for an effective
mean field Hamiltonian which determines this function. The
projected
Hamiltonian is a hermitian operator which acts on this
function, in complete analogy to an ordinary interacting
Hamiltonian, and we may treat it in mean field theory if we
so desire. We accept that the wave functions are enormously
underspecified by this Hamiltonian, but in fact that makes
it more likely, rather than less, that a simple product
will be a fairly good approximation.

The philosophy of this method is analogous to that used by
BCS for superconductivity, and by Laughlin for the
fractional quantum Hall effect: simply guess a wave
function. Is there any better way to solve a non-
perturbative many-body problem?

While the main focus of this paper is on the physical properties 
of the projected wavefunction, we briefly mention what is known about 
its energy as a variational state for the $t$-$J$ model (Hsu, 1990; 
Yokoyama and Ogata, 1996).  At half filling, the projected $d$-wave 
BCS state does remarkably well, with an energy of --0.3199~$J$ per 
bond compared with the best estimate of --0.3346~$J$ 
(Trivedi and Ceperley, 1989).  
Interestingly, projecting the BCS state does just about as well as 
projecting a spin density wave state which has long range order 
(--0.3206~$J$).  This state also has an ordering moment which is much 
too large (0.9).  The best trial state is obtained by combining the 
two, which achieves an energy of --0.3322~$J$ and a staggered 
magnetization of 0.75, which is close to the best numerical 
estimates.
Upon doping, AF co-exists with $d$-wave superconductivity 
up to $x = 0.11$ for $J/t = 0.3$ (Giamarchii and Lhuillier, 1991; 
Himeda and Ogata, 1999; Ogata and Himeda, 2003).  This is in 
disagreement with experiments which show that AF order is destroyed 
beyond 3 to 5\% doping.  However, more recent work which combines 
Gutzwiller projection with a Jastrow factor finds that the energy of 
the $d$-wave superconductor is considerably lowered and
Sorella {\it et al.}\ (2002a) have presented numerical 
evidence that the
ground state of the 2D $t-J$ model has d-wave superconducting long 
range order over a wide doping range;
 see also the work of 
Maier {\it et al.}\ (2000) on the Hubbard model.
This issue is controversial (Zhang {\it et al.}, 1997; Shih {\it et al.}, 1998;
White and Scalapino 1999; Lee {\it et al.}, 2002; Sorella {\it et al.}, 2002b)
and not easy to settle because of technical difficulties with fermion 
simulations.
Nevertheless, the most important point from our perspective is that 
the superconducting
ground state is energetically highly competitive over a broad range of doping,
and thus the variational state whose properties we are describing in 
this paper will
be a good approximation to the ground state of a model close to the 
$t-J$ model.
\\

\noindent
{\bf Mean Field Theory}

In evaluating the energy of these wave functions Zhang {\it et al.} (1988)
used a rough approximation first proposed by Gutzwiller (1963)
which involves assuming complete statistical
independence of the populations on the sites;
see also Vollhardt's (1984) review for a clear explanation. This is not
too bad, since the one-particle states are defined as
momentum eigenstates, but not perfect, as pointed out by
Zhang {\it et al.} (1988) by comparing with Monte Carlo calculations for a
particular case. But in order to understand the results
qualitatively we will follow this simple  procedure here. The
evaluations in Paramekanti {\it et al.} (2001; 2003) are carried out 
without this
approximation.

In the product wave function $\left\vert \Phi\right\rangle $
with the chemical potential fixed so that there are, on
average, $1-x$ electrons per site, with $x$\ the fraction of
holes, the states with 0, 1 and 2 electrons on a given  site
have probabilities
$\left(  1+x\right)  ^{2}/4$, $\left(  1-x^{2}\right)
/2$  and  $\left(   1-x\right)  ^{2}/4$,  respectively.  The
corresponding numbers after projection are $x$, $1-x$ and $0$.
Thus  the relative number of pairs of sites on which a  hole
can  hop  from  one  to the other may be  calculated  to  be
$g_t = 2x/(1+x)$,  while  the relative number of  pairs  of  sites
which can experience
spin exchange is $g_S = 4/(1+x)^{2}$. These are taken to  be  the
renormalization  factors for the kinetic energy and superexchange
terms  in  the  $t-J$ Hamiltonian; that is, the Hubbard Hamiltonian is first
transformed into the $t-J$ Hamiltonian, and then its effect on the
actual product wave-function is estimated in this way. More
accurate estimates could be calculated using Monte-Carlo
methods, and the extra correlated hopping terms could be
included, but we actually doubt if the latter change things
much.

Essentially, in this approximation all terms of  the  nature
of  spin  interactions have a single renormalization factor,
$g_{S}$ while all terms in the kinetic energy are renormalized
by a factor $g_{t}$. The ratio of these is quite large, being about
a factor  of  8 even at 20\%  doping. Thus this method results in an
approximate (or quantitative) spin-charge separation, which is as effective
for experimental purposes (Anderson, 2000) as the
qualitative one of more radical theories. In reality, the
wave function will have some correlations of occupancy, but
these are higher order in $x$ -- in the limit of small $x$
the holes move independently. Also in reality the dispersion
relation may not scale perfectly, but again we do not think
this is a very large effect.

Thus the renormalized Hamiltonian simply takes the form of a
modified $t-J$ Hamiltonian,
\be
H_{eff}=g_{t}T                                             +
g_{S}J\sum\mathbf{S}_{i}\cdot\mathbf{S}_{j}%
\ee
(Again, we ignore the three-site hopping terms). Zhang {\it et al.} (1988)
showed that if we treat this within Hartree-Fock-BCS
approximation, we arrive at a modified BCS gap equation. The
kinetic energy is renormalized downwards, and the
interaction term $\mathbf{S}_{i}\cdot\mathbf{S}_{j}$, which
can be written in the form of four fermion operators
$c^{\dag}c^{\dag}cc$ alike, can be factorized in two ways.
It can be factorized in such a way that it leads to an
anomalous self-energy term of the form,
$J\left\langle
c_{i\uparrow}^{\dag}c_{j\downarrow}^{\dag}\right\rangle
c_{j\downarrow}c_{i\uparrow}+h.c.$, which will lead to a gap;
or it can be factorized in such a way as to give a
Fock exchange self-energy
$\chi_{ij} = \left\langle c_{i\sigma}^{\dag}c_{j\sigma}\right\rangle$,
with $\chi_{\vec  k}$ its Fourier transform, which is  of  nearly
the same form as the kinetic energy, and adds to it.
Exhaustive study of this form of wave function
has led to the conclusion that the optimum  gap
equation solution is a d-wave of symmetry $d_{x^{2}-y^{2}}$
(Kotliar and  Liu, 1988; Suzumura {\it et al.}, 1988; Gros 1988;
Yokoyama and Shiba, 1988; Affleck {\it et al.}, 1988; Zhang {\it et 
al.}, 1988).
The outcome is a pair of coupled
equations, one for the anomalous self-energy and the other
for the effective particle kinetic energy:
\be
\Delta_{\vec k}=\frac{3}{4}g_{S}J\sum_{\vec k}\gamma_{\vec k
-\vec k^{\prime}}\frac{\Delta
_{\vec k^{\prime}}}{2E_{\vec k^{\prime}}}%
\ee
which is an orthodox BCS equation, and%
\be
\chi_{\vec k}=-\frac{3}{4}g_{S}J\sum_{\vec k}\gamma_{\vec k-
\vec k^{\prime}}\frac{\xi_{\vec k^{\prime}%
}}{2E_{\vec k^{\prime}}}.
\ee
Here $\xi_{\vec  k}=g_{t} \varepsilon_{\vec k} - \mu -\chi_{\vec k}$ ,
and $\varepsilon_{\vec k}$ is the band energy, and $\mu$  is
an effective chemical potential, $\gamma_{\vec k}$ is the
Fourier transform  of  the exchange interaction, initially
simply the nearest neighbor result
\be
\gamma_{\vec k}=2\left(  \cos k_{x}+\cos k_{y}\right)
\ee
and  $\mu$ is set to give the right number of electrons
$N_{e}$, which commutes with the projection operator.
$E_{\vec  k}=\sqrt{\xi_{\vec  k}^{2}+\Delta_{\vec  k}^{2}}$,
which
has the same form as in the BCS theory.

\begin{figure}
\begin{center}
\vskip-2mm
\hspace*{0mm}
\epsfig{file=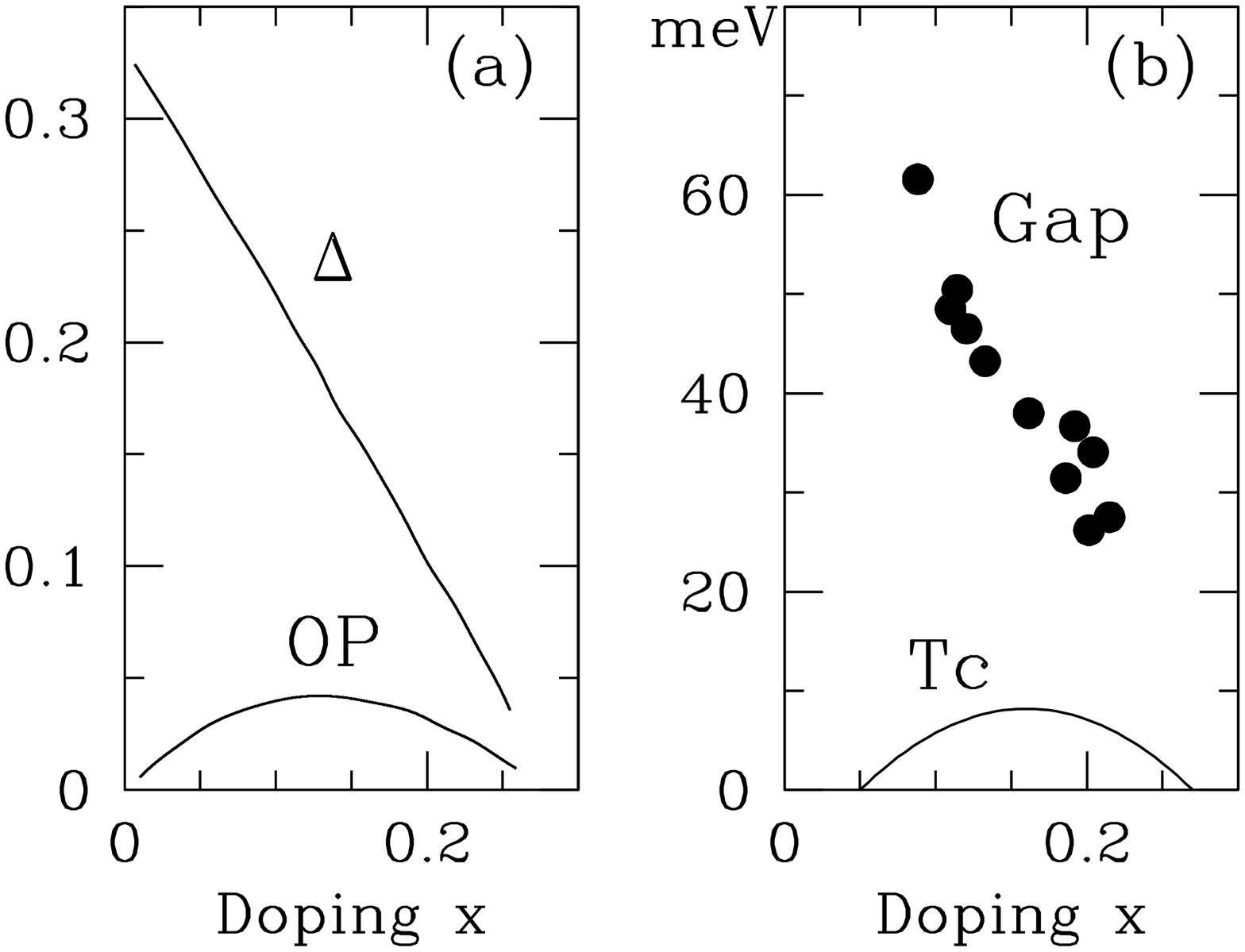,height=4.0in,width=4.0in,angle=0}
\vskip-6mm
\caption{
(a) The amplitude of the (dimensionless) d-wave gap $\Delta$
(called $\tilde{\Delta}$ in Zhang {\it et al.}, (1988)) and the
superconducting order parameter (OP)
as functions of hole doping $x$ in the $t-J$ model for
$J/t=0.2$ calculated in the renormalized
mean field theory of Zhang {\it et al.} (1988).
(b) The spectral gap (in meV) for Bi2212
as measured by ARPES (Campuzano {\it et al.}, 1999)
and $T_c$ as a function of doping.
The $x$ values for the measured $T_c$ were
obtained by using the empirical relation
$T_c/T_c^{\rm max} = 1 - 82.6(x - 0.16)^2$
(Presland {\it et al.}, 1991) with $T_c^{\rm max} = 95$K.
}
\label{fig:gap}
\vskip-6mm
\end{center}
\end{figure}

Zhang {\it et al.} (1988) gave the result of solving these gap equations
in the oversimplified case where only nearest neighbor hopping is
allowed, and we reproduce their figure here as Fig.~2(a).  We
see that $\Delta$, the magnitude of the d-wave symmetry gap,
falls almost linearly with x from a number of order $J$, and
vanishes around $x=0.3$ for $J/t=0.2$. The more realistic
model of Paramekanti {\it et al.} (2001; 2003)
gives a similar result. We presume that this quantity
represents the pseudogap, which is known to vary
experimentally in this way. (A calculation by an entirely
different method (Anderson, 2001) gave the same result.)%

Also plotted on this graph is the physical amplitude of the
order parameter (OP)
$\Delta_{SC}=\left\langle
c_{i\uparrow}c_{j\downarrow}\right\rangle $, which is
supposed to renormalize with $g_{t}$. This is actually true
but the argument is more subtle than that given in Zhang {\it et al.} (1988).
It is necessary to recognize that the two states connected by this
operator contain different numbers of particles.
The simpler argument is
to realize, as was remarked in Paramekanti {\it et al.} (2001; 2003)
that the physically  real
quantity is the off diagonal long range order eigenvalue  of
the density matrix, which is the square root of the product
of $\left\langle c_{i\uparrow
}^{\dag}c_{j\downarrow}^{\dag}c_{i+l\uparrow}c_{j+l\downarrow}
\right\rangle$
for large distance $l$\ which is renormalized by a factor
of $g_{t}^{2}$.
This quantity in this early graph, and in the more accurate
work of Paramekanti {\it et al.} (2001; 2003), bears a striking 
resemblance to the
variation of $T_{c}$ with doping, and was by implication
suggested  to be a measure of $T_{c}$; but it was not until
1997 that the Wen-Lee theory for the renormalization of $T_c$
(to be discussed below) appeared, and it is not quite true that
the order parameter and $T_c$ are identical.

\begin{figure}
\begin{center}
\vskip-2mm
\hspace*{0mm}
\epsfig{file=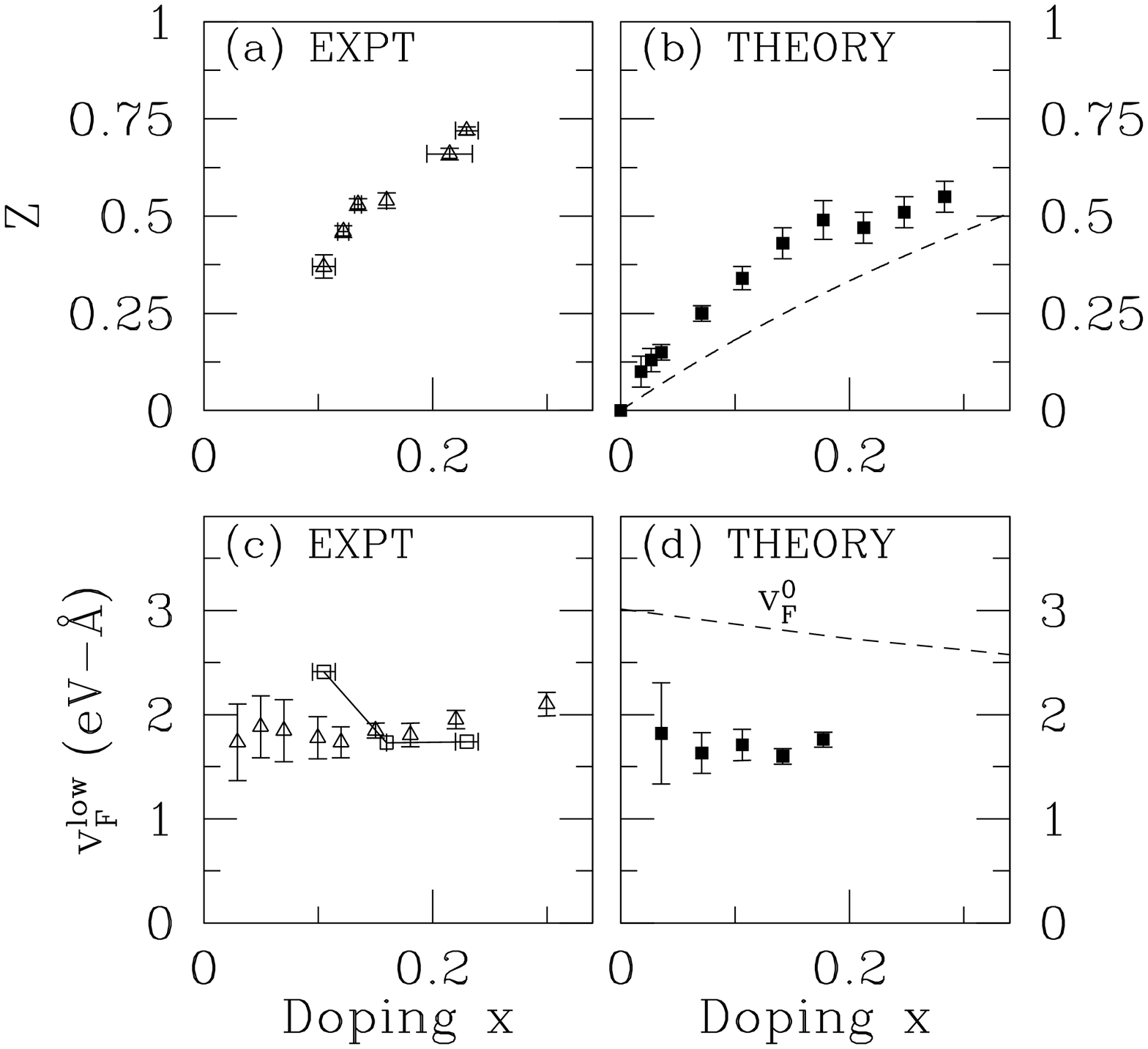,height=4.5in,width=3.5in,angle=0}
\vskip-6mm
\caption{(a): Doping dependence of the nodal quasiparticle
weight $Z$ in Bi2212 extracted from ARPES data (Johnson {\it et al.} 2001) with
$x$ calculated from sample $T_c$ using the
empirical formula of Presland {\it et al.} (1991) with $T^{\rm max}_c = 91K$.
(b): $Z(x)$ predicted from the variational Monte Carlo
calculation of Paramekanti {\it et al.} (2001). The dashed line is the
Gutzwiller approximation result $Z = 2x/(1+x)$.
(c): The low energy nodal Fermi velocity $v_F^{\rm low}$
from ARPES data in Bi2212 (open squares from Johnson {\it et al.} (2001))
and LSCO (open triangles from Zhou {\it et al.} (2003))
is nearly doping independent.
(d): Predicted renormalized $v_F^{\rm low}$
from Paramekanti {\it et al.} (2001) as a function of $x$;
the dashed line is the bare band structure Fermi velocity $v_F^0$.
This figure is adapted from Randeria {\it et al.} (2003).
}
\label{fig:Z}
\vskip-6mm
\end{center}
\end{figure}

Before turning to $T_c$, we briefly mention results on
nodal quasiparticles (``nodons'') obtained from our approach.
These are the important low-lying excitations in the superconducting state
and dominate low temperature thermodynamics, transport and
response functions (Hardy {\it et al.}, 1993; Krishana {\it et al.}, 1995;
Zhang {\it et al.}, 2000; Chiao {\it et al.}, 2000),
in addition to controlling $T_c$ (see below).
The Gutzwiller projected d-wave superconducting ground state
supports sharp nodal quasiparticle excitations (Paramekanti {\it et 
al.}, 2001; 2003)
whose coherent spectral weight
$Z$ goes to zero as $g_t$ but whose Fermi velocity $v_F$ is very 
weakly doping dependent
and remains non-zero as the hole doping $x \to 0$.
These results imply that the real part of the
self-energy $\Sigma^\prime(k,\omega)$ for the gapless nodal quasiparticles has
singular energy and momentum dependences: $Z \sim x$ means that
$\vert\partial \Sigma^\prime / \partial\omega \vert \sim 1/x$ which in turn
implies $\partial \Sigma^\prime / \partial k \sim 1/x$ in order to have a
non-zero nodal $v_F$. These predictions
are in very good agreement with recent ARPES data as shown in Fig.~3, and
in addition also explain the remarkable doping dependence of the 
``high energy''
dispersion of the nodal quasiparticles, above the so-called kink scale
(Lanzara {\it et al.}, 2000), which is found to be dominated by
$\partial \Sigma^\prime / \partial k$ (Randeria {\it et al.}, 2003).

\medskip
\noindent
{\bf Transition Temperature}

It is also a consequence of our theory that the
electromagnetic response function $\rho_{s}$ (the phase
stiffness, or more conventionally $1/\lambda^{2}$, with
$\lambda$ the penetration depth) renormalizes with $g_{t}$,
as does the kinetic energy. In 1997 Lee and Wen (1997) pointed
out that the rate of linear decrease of $\rho_{s}$\  with
temperature, which was the earliest
experimental evidence for d-wave symmetry (Hardy {\it et al.}, 1993),
maintains its magnitude independently of doping.
They argue that the decrease is caused by
the thermal excitation of quasiparticles near the nodes and
is an electromagnetic response function of these
quasiparticles. In the BCS paper it is pointed out that the
electromagnetic response consists of two parts, the
diamagnetic current which is the acceleration in the
field, and the paramagnetic current, which is a
perturbative response of the excited quasiparticles and
exactly cancels the diamagnetic term in the normal state
(Schrieffer, 1964).
The number of these quasiparticles in a d-wave state is only
proportional to $T^{2}$, because the density of states is
only linear in energy. But the amount of decrease of $\rho_s$
per quasiparticle is inversely proportional to its energy,
canceling one factor of $T$.
The key to their argument is the assumption that the current carried by
each quasiparticle is $ev_F$. This is the case in BCS theory, where the
quasiparticle does not carry a definite charge because it is a
superposition of an electron and hole, but each of the partners carries
the same current $ev_F$.  Later it was pointed out by
Millis {\it et al.} (1998) (see also Paramekanti and Randeria (2002))
that there can be a Fermi liquid renormalization of this current to
$\alpha ev_F$ where $\alpha$ is a Fermi liquid parameter inherited from
the normal state. The slope of $\rho_s$ vs $T$ is now proportional to
$\alpha^2$ and we assume that $\alpha$ is of order unity and relatively
insensitive to doping. Thus $\rho_{s}$\ at $T=0$ decreases
proportionally to doping, yet its rate of decrease with temperature
does not vanish with $x$, but instead remains relatively constant.
The decrease of $\rho_s$\ to zero is considered by these
authors to determine $T_{c}$. At $T_{c}$ the system loses
phase coherence, but continues to have an energy gap over
much of the Fermi surface for small $x$. The insensitivity of the linear
$T$ slope in $\rho_s(T)$ to doping was experimentally demonstrated
by Lemberger and co-workers (Boyce {\it et al.}, 2000;
Stajic {\it et al.}, 2003) and verifies our assumption.

As the quasiparticles reduce $\rho_{s}$,
eventually there will develop thermally generated  vortices
(in truly two-dimensional systems like LSCO and Bi2212) and
the actual phase transition takes place as a
Kosterlitz-Thouless (K-T) type of phenomenon (Corson {\it et al.}; 1999)
The notion that a small $\rho_s$ would lead to strong phase fluctuations
which determine $T_c$ was introduced by Emery and Kivelson (1995)
but we must recognize that the $\rho_s$ which controls the K-T
transition is not $\rho_s(T=0)$ but the $\rho_s(T)$ which is greatly
reduced by quasiparticle excitations.  By combining these effects,
the decrease of  $\rho_{s}(T)$  becomes faster than linear, and
eventually infinitely steep. But this happens only quite
near to $T_{c}$, because the K-T $\rho_{s}$ is relatively
low; thus the quasiparticle mechanism gives us a good
estimate of $T_c$, as was pointed out by Lee and Wen, and
fits various empirically proposed relationships
(Uemura {\it et al.}, 1989).
In materials such as YBCO which are more three-dimensional,
the transition  will be more conventional but is still  mediated
by phase fluctuations near $T_{c}$, as of course it is in ordinary
superconductors but not over as broad a critical range.

The Lee-Wen mechanism of $T_c$ described above
is relevant for the underdoped side of the phase diagram
where it offers a natural explanation for $T_c\sim\rho_s(0)$ and holds all the
way up to optimality. On the overdoped side of the phase diagram,
$\rho_s(T)$ continues to be linearly suppressed in
temperature due to thermally excited quasiparticles, but now the stiffness
corresponding to $\rho_s(0)$ is much larger than the energy gap.
Thus superconductivity is lost by gap collapse and $T_c$ would be
expected to scale like the gap for overdoped systems,
as in conventional BCS theory.
\\

\noindent
{\bf Discussion Of Results}

The correspondences between the results of our mean field theory
and the very unusual experimental observations on the high
$T_{c}$ cuprate superconductors are so striking that it is
hard to credit that they have had so little general notice,
especially considering the fact that many of them
constituted predictions made in 1988 before the experimental
situation became clear, sometimes many years before. The d-wave
nature of the energy gap (Kotliar and Liu 1988;
Suzumura {\it et al.} 1988; Gros 1988; Yokoyama and Shiba, 1988)
confirmed only in 1993-94 (Wollmann {\it et al.}, 1993;
Tsuei {\it et al.}, 1994), is the most striking.
The d-wave pairing symmetry was also predicted 
by the ``spin fluctuation theory'' based on a more orthodox structure 
(Bickers, Scalapino and Scalettar, 1987;
Monthoux, Balatsky and Pines, 1991).
This follows earlier predictions of d-wave superconductivity in 
models with strong repulsion in connection with the heavy fermions (Hirsch, 1985; 
Miyake, Schmitt-Rink and Varma, 1986).
We emphasize that our theory, though spin-based, is by construction not
a spin-fluctuation theory, since the latter is based on Fermi liquid theory.
Such a Fermi-liquid based approach may be relevant to the overdoped side of the
cuprate phase diagram, but is unable to deal with the unusual properties
in the vicinity of the Mott insulator.

A second prediction of the RVB approach
is the large energy scale represented by $\Delta$,
which was first observed as a spin gap by NMR at the end of the eighties
(Alloul {\it et al.} 1989; Walstedt and Warren 1990;
Takigawa {\it et al.} 1991). Its significance was only slowly
recognized by the mid-nineties and it has
come to be called the pseudogap. It merges
with the superconducting gap
below $T_{c}$, but is visible in many different kinds of
density of states measurements far above $T_{c}$
(Ding {\it et al.} 1996; Loesser {\it et al.} 1996;
Ch. Renner {\it et al.} 1998). For well
underdoped samples it expunges the Fermi surface in the
anti-nodal direction (Norman {\it et al.}, 1998).
Its value has been studied in
detail by Tallon and Loram (2000), and their numbers are in
striking agreement with the calculations of
Zhang {\it et al.} (1988) or Paramekanti {\it et al.} (2001; 2003),
if we leave aside their claim that it falls to zero in the
midst of the superconducting range.
The pseudogap is often associated
roughly with a temperature scale ``$T^{\ast}$'' below which its effects are
first felt. Of course, in a rigorous sense our mean field theory is a
theory of the superconducting phase at low temperatures, but
the pseudogap appears both in the spectra obtained at low
temperature as well as in the ``mysterious''
pseudogap state above $T_{c}$.

The effects of the renormalization $g_{t}$ on $\rho_{s}$ and
on the Drude weight, which was shown by Sawatzky and coworkers
(Eskes {\it et al.}, 1991; Tajima {\it et al.}, 1990)
to be renormalized with precisely the factor $2x$, 
is a natural consequence of the RVB based theories,
including the mean field theory described here.

One important observation also postdated the
original paper: that the Green's function of the
quasiparticles in the superconducting state
contains a sharp ``coherence peak'' at the quasiparticle
energy on top of a very broad incoherent spectrum, and ARPES
experiments (Feng {\it et al.}, 2000; Ding {\it et al.}, 2001)
have estimated that the amplitude of
that peak is proportional to $2x$. 

One result has not been previously mentioned in the literature. The
renormalization $g_t$ applies to any term in the Hamiltonian which
is a one-electron energy. Therefore matrix elements for
ordinary time-reverse invariant scattering are reduced by a
factor of about $2x$, and their squares, which enter into
such physical effects as the predicted reduction in $T_{c}$,
or into resistivity, are reduced by more than an order  of
magnitude. At the same time the effects of magnetic
scattering are relatively enhanced. Thus the effects of
impurities on high $T_{c}$ superconductivity -- the notorious
contrast of the effects of Zn or Ni substitutions in the plane
relative to non-magnetic doping 
impurities which lie off the plane, (Fukuzumi {\it et al}., 1996)
-- are explained without having any mysterious spin-charge
separation in the formal sense. The same reduction will, on the whole,
apply to the effects of electron-phonon scattering which,
like ordinary impurity scattering, seem to have little
influence on the resistivity.
The electron-phonon interaction, which enters into ordinary BCS 
superconductivity, is renormalized
relative to the spin interaction by the factor $g_t^2/g_S \sim x^2$
and seems unlikely to play a role.

Finally, a word as to the Nernst effect experiments of Ong and coworkers
(Xu {\it et al.}, 2000; Ong and Wang, 2003) which measure
the electric field transverse to an applied thermal gradient in the presence
of a perpendicular magnetic field. The Nernst signal is expected to be
dominated by the motion of vortices, and the results on two
dimensional materials are very consistent with expectations
for a generalization of the Kosterlitz-Thouless type of
transition.  What is seen is a Nernst signal at and below
$T_{c}$ varying at low magnetic field $B$ as $B\ln B$
(Ong and Wang, 2003) indicating that the underlying
$\rho_{s}$ of the effective Ginsburg-Landau free energy does
not vanish at $T_{c}$; the $\ln B$\ variation, giving an
infinite slope, follows from thermal proliferation of large
vortices whose energies vary as $\rho_{s}\ln B$. As $B$
is increased, however, the signal does not  drop to zero
until a very large $B$\ is reached, indicating a retention
of phase stiffness at short length scales long after
superconducting long-range order has disappeared. We believe
that this is a natural and probably calculable effect. But
with increasing temperature the Nernst
effect disappears well below $T^*$, at least for low fields.
In this region we are well out of the region of
applicability of mean field theory, and expect very large 
fluctuation effects for which we have no controlled theory.

An additional experimental phenomenon which, we think,
supports the essential validity of a projected wave
function is the particle-hole asymmetry of the
tunneling conductance as a function of voltage.
We will discuss single-particle excited states and tunneling
asymmetry in a forthcoming paper (Anderson {\it et al.}, 2003)
\\

\noindent
{\bf Conclusion}

In broad outline, our basic assumptions as to the physics of
the cuprates, together with a mean field theory which is
little less manageable than BCS theory, seem to give a remarkably
complete picture of the unusual nature of the superconducting state. 
The RVB state is still a pairing state 
between electrons.  It has its genesis in the BCS state and is 
smoothly connected to it, a fact which is made clear in the recent 
studies of a partially projected BCS state (Laughlin, 2002; Zhang, 
2003).  Furthermore, its low lying excitations are well defined 
quasiparticles which dominate the low temperature physics.  Thus the 
RVB state is in some ways rather conventional.  What is unusual is 
the reduction of the superfluid density and the quasiparticle 
spectral weight.  With increasing degrees of projection, the state 
evolves from pairing of quasiparticles to one which is better 
understood as a spin singlet formation with coherent hole motion. 
This evolution has the following dramatic consequence.  The BCS 
pairing is driven by a gain in the attractive potential at the 
expense of kinetic energy, since the energy gap smears out the Fermi 
occupation $n(\vec k)$.  With projection, $n(\vec k)$ is already 
strongly smeared in the non-Fermi liquid normal state, and 
superconductivity is instead stabilized by a gain in kinetic energy 
due to coherent hole motion.  This picture has been verified by 
experiments which monitor the kinetic energy via the optical sum rule 
(van der Marel {\it et al}., 2003)

Why then is the subject so
controversial?  Aside from purely socio-political reasons,
there is a real difficulty: the proliferation of nearby
alternative states of different symmetry. 
Here we mention a number 
of possibilities that are actively being considered.  One important 
issue is the evolution to the antiferromagnet at very low doping.  On 
general principles (Baskaran, 2000; Anderson and Baskaran, 2001), 
mesoscopically inhomogeneous states (``stripes'') are likely to be 
stable at low doping on some scale.  They show up in some numerical 
calculations (White and Scalapino, 1999) and a few of the cuprates 
show indications of them as static (Tranquada {\it et al}., 1995) or 
dynamical excitations (see Stock {\it et al}., 2004). While static 
stripes are undoubtedly detrimental to superconductivity, there have 
been arguments that dynamical stripes may be the source of pairing 
(see Carlson {\it et al}., 2004).  We note that in this scenario, the 
pairing originates from the ladder structure of the hole-free part of 
the stripe which also has its origin in RVB physics.  Given the 
success of the uniform projected wavefunction, we find these more 
complex scenarios neither necessary nor sufficient for the 
intermediate doping range.

A second class of competing states has 
its origin in the $SU$(2) gauge symmetry first identified for the 
projected wavefunction at half-filling. The states of an
undoped RVB, or in fact any state of the Mott insulator, can
be  represented  by an enormous number  of wave functions
before projection; in fact, as pointed out by
Affleck {\it et al.} (1988) (see also Anderson (1987b) and
Zhang {\it et al.} (1988)), it has an SU(2) gauge symmetry.
In the undoped state, with exactly one electron per site, the presence
of an up spin is equivalent to the absence of a down spin and vice versa,
thus permitting independent $SU$(2) rotations at each site.
This degeneracy in the representation
of the wave function does not imply any true degeneracy, it is merely
the consequence of our using an underdetermined representation.

When we add holes, this
gauge freedom gradually  becomes physical, which we
experience as the development of a stiffness to phase
fluctuations which grows from zero proportionally to $x$.
The fluctuations can actually take place in a larger space
of gauge degrees of freedom which we can represent in terms
of staggered flux phases, etc.
(Affleck and Marston, 1988) as possible Hartree-Fock
states, but we expect that these are higher energy
than the superconducting state for the interesting values of $x$.
However, the energy difference is small for small $x$, and Wen 
and Lee (1996) have proposed that in the underdoped region $SU$(2) 
rotations which connect fluctuations of staggered flux states and 
$d$-wave superconductivity may play a role in explaining the 
pseudogap phenomenon.  Remarkably, orbital current correlations which 
decay rather slowly as power law have been seen in projected $d$-wave 
wavefunctions (Ivanov {\it et al}., 2000).  These fluctuations are 
very natural in the $SU$(2) gauge theory but are otherwise 
unexpected.  In a related development, a static orbital current 
state, called $d$-density wave, has been proposed to describe the 
pseudogap on phenomenological grounds (Chakravarty {\it et al}., 
2001).

In this review we have focused our attention on the ground 
state and low-lying excitations in the underdoped region.  Due to the 
multitude of competing states mentioned above, much work remains 
before a full understanding of the pseudogap is achieved.  The 
situation becomes even worse for doping to the right of the $T^\ast$ 
crossover line, commonly called the ``strange metal'' phase.  Here 
one sees highly anomalous transport properties such as the linear 
resistivity which played such an important role in early thinking. 
While the RVB theory leads naturally to a crossover from pseudogap to 
strange metal and to Fermi liquid as one increases the doping at a 
temperature above the  optimal $T_c$, the ideas presented here are no 
help in understanding the breakdown of Fermi liquid behavior in the 
strange metal. Instead of a smooth crossover, many workers ascribe 
the anomalous behavior to a quantum critical point which lies in the 
middle of the superconducting dome (Varma, 1997;  Tallon and Loram, 
2000; Varma, 2003).  We simply remark that the quantum critical 
point, if it exists, is different from any previous examples in that 
there is no sign of a diverging correlation length scale in any 
physical observable, and it is difficult to draw lessons from past 
experience even phenomenologically.

Finally, what about phonons? Of course there is some coupling to
optical phonons, which will influence both the phonons
themselves -- an influence which will change sign with the
phonon wave vector $Q$, because of coherence factors --
and the dispersion of quasiparticles. But as remarked,
the net effect of an optical phonon on d-wave
superconductivity will tend to cancel out over the Brillouin
zone. It certainly will not play a controlling role in a system so
dominated by Coulomb repulsion. In any case, phonon effects
on electron self-energies will tend to be renormalized
downwards by the square of $g_{t}$, as we pointed out before.

We close by remarking that great strides have been made in the 
discovery of unconventional superconductors since 1986. Today, 
non-$s$-wave pairing states are almost commonplace in heavy fermions, 
organic superconductors, and in transition metal oxides. Even time 
reversal symmetry is not sacrosanct (see the review on Sr$_2$RuO$_4$ 
by MacKenzie and Maeno, 2003).  The discovery of high $T_c$ has 
opened our eyes to the possibility that superconductivity is an 
excellent choice as the ground state of a strongly correlated system. 
This may be the most important message to be learned from this 
remarkable discovery.
\\

\noindent
{\bf Acknowledgments}

We would like to acknowledge collaborations and interactions with
many people over the years, with special thanks to G. Baskaran,
C. Gros, R. Joynt, A. Paramekanti and X.-G. Wen.
PAL would like to acknowledge NSF DMR-0201069. MR would like to
thank the Indian DST for support under the Swarnajayanti scheme,
and the Princeton MRSEC NSF DMR-0213706 for supporting
his visits to Princeton University.

%

\bigskip
\bigskip
\noindent
{\bf Acknowledgments}
\bigskip

\medskip
\noindent
Affleck, I., and J. B. Marston, 1988, Phys. Rev. B37, 3774.

\medskip
\noindent
Affleck, I., Z. Zou, T. Hsu, and P. W. Anderson, 1988, Phys. Rev. B38, 745.

\medskip
\noindent
Alloul, H., T. Ohno and P. Mendels, 1989, Phys. Rev. Lett. 63, 1700.

\medskip
\noindent
Anderson, P. W., 1959, Phys. Rev. 115, 2.

\medskip
\noindent
Anderson, P. W., 1973, Mater. Res. Bull. 8, 153.

\medskip
\noindent
Anderson, P. W., 1987a, Science 237, 1196.

\medskip
\noindent
Anderson, P. W., 1987b, in ``Frontiers and Borderlines in
Many Particle Physics'', Proceedings of the Enrico Fermi
International School of Physics, Varenna (North Holand).

\medskip
\noindent
Anderson, P. W., 1997, \textquotedblleft  The Theory of High Tc
Superconductivity\textquotedblright, Ch.~1, (Princeton University Press).

\medskip
\noindent
Anderson, P. W., 2000 Science 288, 480.

\medskip
\noindent
Anderson, P. W., 2001, cond-mat/0108522.

\medskip
\noindent
Anderson, P. W., 2002 Physica Scripta T102, 10.

\medskip
\noindent
Anderson, P. W., and G. Baskaran, 2001, unpublished.

\medskip
\noindent
Anderson, P. W., G. Baskaran, Z. Zou, and T. Hsu, 1987,
Phys. Rev. Lett. 58, 2790.

\medskip
\noindent
Anderson, P. W., and P. Morel, 1961, Phys. Rev. 123, 1911.

\medskip
\noindent
Anderson, P. W., {\it et al}., 2003, unpublished.

\medskip
\noindent
Baskaran, G., 2000, Mod. Phys. Lett. B14, 377.


\medskip
\noindent
Baskaran, G., and P. W. Anderson, 1988, Phys. Rev. B 37, 580.

\medskip
\noindent
Baskaran, G., Z. Zou, and P. W. Anderson, 1987, Sol. St. Comm. 63, 973.


\medskip
\noindent
Bednorz, J. G., and K. A. M\"uller, 1986, Z. Phys. B64, 189.

\medskip
\noindent
Bethe, H. A., 1931, Z. Phys. 71, 205.

\medskip
\noindent
Bickers, N. E., D. J. Scalapino, and R. T. Scalettar, 1987, Int. J. 
Mod. Phys. B1, 687.

\medskip
\noindent
Boyce, B. R., J. A. Skinta, and T. Lemberger, 2000, Physica C 341-348, 561.

\medskip
\noindent
Campuzano, J.-C. {\it et al.},
1999, Phys. Rev. Lett. 83, 3709.

\medskip
\noindent
Carlson, E.W., V.J. Emery, S.A. Kivelson, and D. Orgad, 
2002, cond-mat/0206217, to appear in ``The Physics of Conventional 
and Unconventional Superconductors,'' ed. K. Benneman and J.B. 
Ketterson (Springer-Verlag).

\medskip
\noindent
Chakravarty, S., 
R.B. Laughlin, D. Morr, and C. Nayak, 2001, Phys. Rev. B63, 94503.

\medskip
\noindent
Chiao, M., R.W. Hill, C. Lupien, L. Taillefer, P. Lambert, R. Gagnon, and
P. Fournier, 2000, Phys. Rev. B. 62, 3554.

\medskip
\noindent
Corson, J., R. Mallozzi, J. Orenstein, J. N. Eckstein, and I. Bozovic,
1999, Nature 398, 221.

\medskip
\noindent
Ding, H., {\it et al.}, 1996, Nature 382, 51.

\medskip
\noindent
Ding, H., J. R. Engelbrecht, Z. Wang, J. C. Campuzano, S. C. Wang,
H. B. Yang, R. Rogan, T. Takahashi, K. Kadowaki, and D. G. Hinks,
2001, Phys. Rev. Lett. 87, 227001.

\medskip
\noindent
Emery, V. J., and S. A. Kivelson, 1995 Nature 374, 434.

\medskip
\noindent
Eskes, H., M. B. J. Meinders, and G. Sawatzky, 1991,
Phys.  Rev. Lett. 67, 1035

\medskip
\noindent
Fazekas, P., and P. W. Anderson, 1974, Phil. Mag. 30, 432.

\medskip
\noindent
Feng, D.L., D.H. Lu, K.M. Shen, C. Kim, H. Eisaki, A. Damascelli,
R. Yoshizaki, J.-I. Shimoyama, K. Kishio, G.D. Gu, S. Oh, A. Andrus,
J. O'Donnell, J.N. Eckstein, and Z.-X. Shen, 2000 Science 289, 277.


\medskip
\noindent
Fukuyama, H., 1992, Prog. Theor. Phys. 108, 287.

\medskip
\noindent
Fukuzumi, Y., K. Mizukashi, K. Takenaka, and S. 
Uchida, 1996, Phys. Rev. Lett. 76, 684.

\medskip
\noindent
Giamarchi, T., and C. Lhuillier, 1991, Phys. Rev. B43, 12943.

\medskip
\noindent
Gros, C., 1988 Phys. Rev. B38, 931.

\medskip
\noindent
Gros, C., 1989, Annals of Phys. 189, 53.

\medskip
\noindent
Gros, C., R. Joynt, and T. M. Rice, 1986, Phys. Rev. B36, 381.

\medskip
\noindent
Gutzwiller, M. C., 1963 Phys. Rev. Lett. 10, 159.

\medskip
\noindent
Hardy, W. N., D. A. Bonn, D. C. Morgan, R. Liang, and K. Zhang,
1993, Phys. Rev. Lett. 70, 3999.

\medskip
\noindent
Himeda, A. and M. Ogata, 1999, Phys. Rev. B60, 9935.

\medskip
\noindent
Hirsch, J. E., 1985, Phys. Rev. Lett. 54, 1317.

\medskip
\noindent
Hsu, T. C., 1990, Phys. Rev. B41, 11379.

\medskip
\noindent
Ioffe, L., and A. I. Larkin, 1989, Phys. Rev. B39, 8988.

\medskip
\noindent
Ivanov, D.A., P.A. Lee, and X.G. Wen, 2000, 
Phys. Rev. Lett. 84, 3958.

\medskip
\noindent
Johnson, P. D., {\it et al}.,
2001 Phys. Rev. Lett. 87, 177007

\medskip
\noindent
Kivelson, S. A., D. S. Rokhsar, and J. P. Sethna,
1987, Phys. Rev. B 38, 8865.

\medskip
\noindent
Kohn, W., 1964, Phys. Rev. 133, A171.

\medskip
\noindent
Kohn, W., and J. M. Luttinger, 1965, Phys. Rev. Lett. 15, 524.

\medskip
\noindent
Kotliar, G., and J. Liu, 1988, Phys. Rev. B38, 5142.

\medskip
\noindent
Krishana, K., J. M. Harris, and N.P. Ong, 1995 Phys. Rev. Lett. 75, 3529.

\medskip
\noindent
Lanzara, A., {\it et al.}, 2001, Nature 412, 510.

\medskip
\noindent
Laughlin, R. B., 2002, cond-mat/0209269.

\medskip
\noindent
Lee, P. A., 1989, Phys. Rev. Let. 63, 680.

\medskip
\noindent
Lee, P. A., and X. G. Wen, 1997, Phys. Rev. Lett 78, 4111.

\medskip
\noindent
Lee, P. A., and X. G. Wen, 2001, Phys. Rev. B 63, 224517.

\medskip
\noindent
Lee, T. K., C. T. Shih, Y. C. Chen, and H. Q. Lin, 2002,
Phys. Rev. Lett. 89, 279702.

\medskip
\noindent
Loesser, A. G., {\it et al.}, 1996, Science 273, 325.

\medskip
\noindent
MacKenzie, A.P., and Y. Maeno, 2003, Rev. Mod. 
Phys. 75, 657.

\medskip
\noindent
Maier, Th., M. Jarrell, Th. Pruschke, and J. Keller, 2000,
Phys. Rev. Lett. 85, 1524.

\medskip
\noindent
Monthoux, P., A. Balatsky, and D. Pines, 1991, Phys. Rev. Lett. 67, 3449.

\medskip
\noindent
Millis, A. J., S. M. Girvin, L. B. Ioffe, and A. I. Larkin,
1998 J. Phys. Chem. Solids 59, 1742.

\medskip
\noindent
Miyake, K., S. Schmitt-Rink and C. M. Varma, 1986, Phys. Rev. B34, 6654.

\medskip
\noindent
Nagaosa, N., and P. A. Lee, 1990, Phys. Rev. Lett. 64, 2450.


\medskip
\noindent
Norman, M. R., H. Ding, M. Randeria, J. C. Campuzano,
T. Yokoya, T. Takeuchi, T. Takahashi, T. Mochiku, K. Kadowaki,
P. Guptasarma, and D. G. Hinks, 1998, Nature 392, 157.

\medskip
\noindent
Ogata, M. and A. Himeda, 2003, J. Phys. Soc. Jpn. 72, 374; (cond-mat/0003465).

\medskip
\noindent
Ong, N. P., and Y. Wang, 2003, cond-mat/0306399, and private communication.

\medskip
\noindent
Paramekanti, A., and M. Randeria, 2002, Phys. Rev. B 66, 214517.


\medskip
\noindent
Paramekanti, A., M. Randeria, and N. Trivedi, 2001,
Phys. Rev. Lett. 87, 217002.

\medskip
\noindent
Paramekanti, A., M. Randeria, and N. Trivedi, 2003,
cond-mat/0305611 (to appear in Phys. Rev. B).


\medskip
\noindent
Presland, M. R., J. R. Tallon, R. G. Buckley, R. S. Liu, and N. E. Flower,
1991, Physica C 176, 95.

\medskip
\noindent
Randeria, M., A. Paramekanti, and N. Trivedi, 2003, cond-mat/0307217
(to appear in Phys. Rev. B).

\medskip
\noindent
Ch. Renner, B. Revaz, J.-Y. Genoud, K. Kadowaki, and O. Fischer, 1998,
Phys. Rev. Lett. 80, 149.


\medskip
\noindent
Ruckenstein, A., P. Hirschfeld, and J. Appel, 1987, Phys. Rev. B 36, 857.

\medskip
\noindent
Schrieffer, J. R., 1964,
\textquotedblleft  Theory of Superconductivity\textquotedblright, (Benjamin).

\medskip
\noindent
Shen, Z.X., A. Lanzara, S. Ishihara, and N. 
Nagaosa, 2002, Phil. Mag. B82, 1349.

\medskip
\noindent
Shih, C. T., Y. C. Chen, H. Q. Lin, and T. K. Lee,
1998, Phys. Rev. Lett. 81, 1294.

\medskip
\noindent
Shraiman, B. and E. Siggia, 1989, Phys. Rev. Lett. 62, 1564.

\medskip
\noindent
Sorella, S., G. B. Martins, F. Becca, C. Gazza, L. Capriotti,
A. Parola, and E. Dagotto, 2002a, Phys. Rev. Lett. 88, 117002.

\medskip
\noindent
Sorella, S., {\it et al.}, 2002b, Phys. Rev. Lett. 89, 279703;

\medskip
\noindent
Stajic, J., {\it et al}., 2003 Phys. Rev. B 68, 024520.

\medskip
\noindent
Stock, C. {\it et al}., 2004, Phys. Rev. B69, 
014502.

\medskip
\noindent
Suzumura, Y., Y. Hasegawa, and H. Fukuyama, 1988, J. Phys. Soc. Jpn. 57,2768.

\medskip
\noindent
S. Tajima {\it et al.}, 1989, Proc. Tsukuba Conf. II, T. Shiguro and 
K. Kajimura eds.,
(Springer, 1990).

\medskip
\noindent
Tallon, J. L., and J. W. Loram, 2000, Physica C 349, 53.

\medskip
\noindent
Takigawa, M., A. P. Reyes, P. C. Hammel, J. D. Thompson, R. H. Heffner,
Z. Fisk and K.C. Ott, 1991, Phys. Rev. B 43, 247.

\medskip
\noindent
Tranquada, J. {\it et al}., 1995, Nature 375, 561.

\medskip
\noindent
Trivedi, N., and D. M. Ceperley, 1989, Phys. Rev. B40, 2737.

\medskip
\noindent
Tsuei, C.C., J. R. Kirtley, C. C. Chi, L. S. Yujahnes, A. Gutpa, T. Shaw,
J. Z. Sun, and M. B. Ketchen, 1994, Phys. Rev. Lett. 73, 593.

\medskip
\noindent
Uemura, Y.J., {\it et al}., 1989, Phys. Rev. Lett. 62, 2317.

\medskip
\noindent
van der Marel, D., H.T.A. Molegraaf, C. 
Presura, and I. Santoso, 2003, 
cond-mat/0302169.

\medskip
\noindent
Varma, C.M., 1997, Phys. Rev. 
B55, 14554.

\medskip
\noindent
Varma, C.M., 2003, cond-mat/0312385.

\medskip
\noindent
Vollhardt, D., 1984, Rev. Mod. Phys. 56, 99.

\medskip
\noindent
Walstedt, R. E., and W. W. Warren, 1990, Science 248, 82.

\medskip
\noindent
Wen, X. G., and P. A. Lee, 1996, Phys. Rev. Lett. 76, 503.

\medskip
\noindent
Weng, Z. Y., D. N. Sheng, and C. S. Ting, 1998, Phys. Rev. Lett. 80, 5401.

\medskip
\noindent
White, S. R., and D. J. Scalapino, 1999, Phys. Rev. B 60, 753.

\medskip
\noindent
Wiegmann, P. B., 1988, Phys. Rev. Lett. 60, 821.

\medskip
\noindent
Wollman, D. A., D. J. vanHarlingen, W. C. Lee, D. M. Ginsberg,
and A. J. Leggett, 1993, Phys. Rev. Lett. 71, 213.

\medskip
\noindent
Xu, Z. A., N. P. Ong, Y. Wang, T. Kageshita, and S. Uchida, 2000, 
Nature 406, 486.

\medskip
\noindent
Yokoyama, H., and H. Shiba, 1988 J. Phys. Soc. Jpn. 57, 2482.

\medskip
\noindent
Yokoyama, H., and M. Ogata, 1996, J. Phys. Soc, 
Jpn, 65, 3615.

\medskip
\noindent
Zhang, F. C., 2003, Phys. Rev. Lett. 90, 207002.

\medskip
\noindent
Zhang, F. C., C. Gros, T. M. Rice, and H. Shiba, 1988,
Supercond. Sci. Tech. 1, 36.

\medskip
\noindent
Zhang, F. C., and T. M. Rice, 1988, Phys. Rev. B37, 3759.

\medskip
\noindent
Zhang, S., J. Carlson, and J. Gubernatis, 1997, Phys. Rev. Lett. 78, 4486.

\medskip
\noindent
Zhang, Y., N. P. Ong, Z. A. Xu, K. Krishana, R. Gagnon, and L. Taillefer,
2000, Phys. Rev. Lett. 84, 2219.

\medskip
\noindent
Zhou, X. J., {\it et al}.,
2003, Nature 423, 398.

\medskip
\noindent
Zou, Z. and P. W. Anderson, 1988, Phys. Rev. B 37, 627.


\end{document}